\documentclass[onecolumn,preprintnumbers,amsmath,amssymb]{revtex4}
\usepackage{graphicx}% Include figure files
\usepackage{dcolumn}% Align table columns on decimal point
\usepackage{bm}% bold math
\usepackage{times}

\begin{document}

\title{An Artificially Lattice Mismatched Graphene/Metal Interface: Graphene/Ni/Ir(111)}

\author{Daniela Pacil\'e$^{1, 2}$\footnote{Email address: daniela.pacile@fis.unical.it}}
\author{Philipp Leicht$^{3}$}
\author{Marco Papagno$^{1, 2}$}
\author{Polina M. Sheverdyaeva$^{2}$}
\author{Paolo Moras$^{2}$}
\author{Carlo Carbone $^{2}$}
\author{Konstantin Krausert$^{3}$}
\author{Lukas Zielke$^{3}$}
\author{Mikhail Fonin$^{3}$}
\author{Yuriy S. Dedkov$^{4,}$\footnote{Present address: SPECS Surface Nano Analysis GmbH, Berlin, Germany}}
\author{Florian Mittendorfer$^{5}$}
\author{J\"org Doppler$^{5}$}
\author{Andreas Garhofer$^{5}$}
\author{Josef Redinger$^{5}$}

\newcommand{\degC}{$^{\circ}\mathrm{C}$ }

\affiliation{
$^{1}$~\mbox{Dipartimento di Fisica, Universit$\grave{a}$ della Calabria, 87036 Arcavacata di Rende (CS), Italy} \\
$^{2}$~\mbox{Istituto di Struttura della Materia, Consiglio Nazionale delle Ricerche, Trieste, Italy}\\
$^{3}$~\mbox{Fachbereich Physik, Universit\"at Konstanz, 78457 Konstanz, Germany}\\
$^{4}$~\mbox{Fritz-Haber-Institut der Max-Planck Gesellschaft, Faradayweg 4-6, 14159 Berlin, Germany}\\
$^{5}$~\mbox{Institute for Applied Physics and Center for Computational Materials Science, Vienna University of Technology, 1040 Vienna, Austria}\\
}

\date{\today}% It is always \today, today,
             %  but any date may be explicitly specified

\begin{abstract}
{We report the structural and electronic properties of an artificial graphene/Ni(111) system obtained by the intercalation of a monoatomic layer of Ni in graphene/Ir(111). Upon intercalation, Ni grows epitaxially on Ir(111), resulting in a lattice mismatched graphene/Ni system. By performing Scanning Tunneling Microscopy (STM) measurements and Density Functional Theory (DFT) calculations,  we show that the intercalated Ni layer leads to a pronounced buckling of the graphene film. At the same time an enhanced interaction is  measured by Angle-Resolved Photo-Emission Spectroscopy (ARPES), showing  a clear transition from  a nearly-undisturbed to a strongly-hybridized graphene $\pi$-band. A comparison of the intercalation-like graphene system with flat graphene on bulk Ni(111), and mildly corrugated graphene on Ir(111), allows to disentangle the two key properties which lead to the observed increased interaction,  namely  lattice matching and electronic interaction.  Although the latter determines the strength of the hybridization, we find an important influence of the local  carbon configuration resulting from the lattice mismatch.}
\end{abstract}

% PACS, the Physics and Astronomy
                             % Classification Scheme.
%\keywords{KEYWORDS: graphene, transition metals, moir\'e structures}%Use showkeys class option if keyword
                              %display desired

\maketitle
\section{Introduction}
Metal supported graphene has received renewed interest as it provides a model-system for studying graphene modifications on well-defined large area samples. Recent photoemission and Scanning Tunneling Microscopy (STM) studies have shown that the carrier mobility, chirality, and band gap can be tailored by a periodic perturbation potential \cite{Rusponi}, doping \cite{Nagashima, Papagno}, intercalation \cite{Varykhalov}, and hybridization with the supporting substrate \cite{Batzill}. Although there are countless studies on graphene grown on transition metals \cite{Wintterlin,Batzill,Dedkov:2012}, the vastly differing interaction of graphene (G) with transition metals (Me) is not fully understood on a basic level~\cite{Voloshina:2012, Kozlov}. As an important contribution to the interaction stemms from non-local (van-der-Waals like) interactions, the variability of a G film with the supporting metal is only partially explained by the so-called $\textit{d}$ band model \cite{Hammer}. That Pt and Ir interact more weakly with G than Ni or Co is not surprising according to this model, but changes from one metal to the next one in the periodic table are expected to be more gradual. Instead, for neighboring elements, like Pd and Pt, and Rh and Ir, the G-Me interaction seems to abruptly switch from $\textit{strong}$ to $\textit{weak}$ \cite{Batzill, Wintterlin}. According to the experimental findings, the G-Me interaction has been partitioned into these two main categories, where from an electronic structure point of view, a $\textit{strong}$ or a $\textit{weak}$ interaction means a \textit{perturbed} or an almost \textit{unperturbed} graphene $\pi$-band at the K-point of the Brillouin zone. On the other hand, the term $\textit{strong}$ appears inappropriate if intended for chemisorption between graphene and the underlaying metal. Indeed, for Ni(111), which is considered one of the metals belonging to the $\textit{strong}$ category, only a moderate adsorption energy of 67 meV per carbon atom has been recently evaluated on the basis of high-level many-body calculations (and up to 160 meV/C using semi-empirical forcefield corrections), which still is in the range of typical physisorption systems \cite{Mittendorfer,Dedkov3,Adamska}. Therefore, it should be noted that the strong interaction mainly implies a strong hybridization between the graphene $\pi$ states and the substrate, but is not necessarily reflected in the adsorption energies. A related question is a possible correlation between the G-Me lattice mismatch and the strength of interaction. Several metals belonging to the $\textit{strong}$ (Rh, Ru) and $\textit{weak}$ (Ir, Pt, Cu) categories form moir\'e structures, comprising different interactions with the G layer. 

Among several transition metals, Ni(111) has been most studied as substrate material for G-Me interface, both by  theory and experiment. 
The close lattice match between G and Ni allows the growth of a commensurate (1$\times$1) graphene overlayer, with carbon atoms at atop and fcc-hollow sites, separated from the substrate 
by 2.11 $\mbox\AA$ and 2.16 $\mbox\AA$, respectively \cite{Gamo}. 
Angle-resolved photoemission (ARPES) data show a pronounced energy gap at the K-point of the Brillouin zone between $\pi$ and $\pi{^\ast}$ \cite{Dedkov1}, as a result of broken symmetry for the two carbon sublattices accompained by strong hybridization between Ni 3d and graphene $\pi$ states. 

Here we report a new G-Ni(111) system, obtained by the intercalation of a single epitaxial layer of Ni in graphene/Ir(111). For this system, the lattice mismatch between graphene and the Ni layer is increased. Although the epitaxial growth of Ni on Ir(111) will lead to additional electronic effects, such as a narrowing of the Ni $\textit{d}$ band, the local chemical electronic environment is still similar enough to allow for a comparision with G/Ni(111). The intercalation leads to a {\it locally} enhanced interaction, resulting in a strong corrugation of the graphene layer. We investigated the artificially mismatched G/Ni by  STM, Density Functional Theory (DFT) including van der Waas contributions (vdW-DF) and ARPES, providing a wide characterization of electronic and structural properties. The comparison between G/Ni/Ir(111) and G/Ni(111) allows to rank the influence of two important factors, lattice mismatch and chemical interaction,  affecting the G-Me adsorption mechanism.

\section{Methods}
The presented studies were performed in two different experimental chambers under identical experimental conditions, allowing for a reproducible sample preparation. 
Photoemission experiments were carried out at the VUV-Photoemission beam line of the Elettra synchrotron radiation facility (Trieste, Italy), using a Scienta R4000 electron energy analyzer at a base pressure of $5\times 10^{-11}$\,mbar. Angle-resolved photoemission spectra were collected at room temperature (RT) using a photon energy of 80 eV, with total energy resolution of 100~meV and angular resolution of 0.1$^{\circ}$. 

STM experiments were carried in ultra-high vacuum (UHV) system (base pressure $5\times 10^{-11}$\,mbar) equipped with an Omicron variable temperature scanning tunneling microscope. 
All STM measurements were performed in the constant-current-mode at RT using electrochemically etched polycrystalline tungsten tips cleaned in UHV by flash-annealing. 
The sign of the bias voltage corresponds to the voltage applied to the sample.  Tunneling current and voltage are labeled $I_{T}$ and $U_{T}$, respectively. 

Experimentally, the G/Ir(111) system was prepared by the procedure described in \cite{Rusponi}. Intercalation of Ni underneath a graphene layer was performed via annealing of the pre-deposited film in the temperature range of 670-800 K. Starting from the sub-monolayer regime, the Ni coverage was estimated on bare Ir(111) by measuring the intensity ratio of Ni-3p and Ir-4f core levels.

The spin-polarized DFT calculations were performed with the Vienna Ab-initio Simulation Package (VASP) \cite{vasp1,vasp2}, using PAW potentials \cite{paw1,paw2} and an energy-cutoff of 400 eV. 
As GGA exchange-correlation functionals tend to severely underbind the adsorption of graphene on Ni(111) \cite{Mittendorfer}, the calculations were performed using van der Waals DFT (vdW-DF)  
with the opt86b functional \cite{vdW,vdW2}. In the calculations, a ($10 \times 10$) graphene sheet was adsorbed on a   ($9 \times 9$) Ir(111) substrate with a lattice constant of 2.735 \AA, 
consisting of a three layer slab and an additional intercalated epitaxial Ni layer.  
A $\Gamma$-centered $3\times3\times1$ k-point mesh was used to relax
the structures keeping only the two bottom-most layers fixed. The C 1s core level shifts were calculated in the initial state approximation. For the graphical visualisation, the resulting total core level spectra are displayed as a sum over Gaussian functions with a standard deviation of 0.25. The STM simulations were performed using the Tersoff-Hamann approximation \cite{TersoffHamann} using the integrated charge density between E$_F$ and E$_F$ + 0.2\,eV.

 \begin{figure}
\includegraphics[width=3.2in]{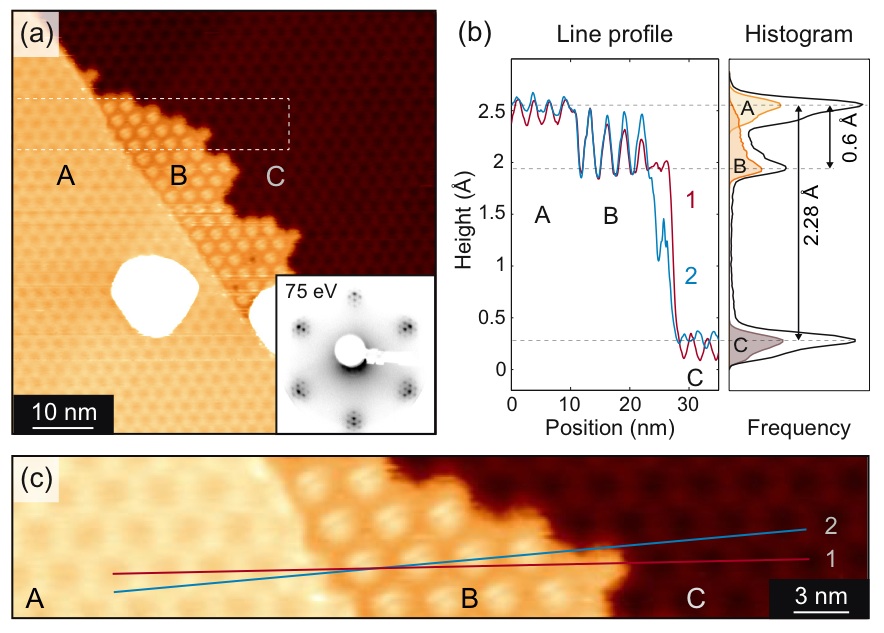}
\caption{\label{fig_stm_0:epsart} (color online) (a) Topographic STM overview showing the morphology of graphene with a partially intercalated Ni sub-monolayer. Ni accumulates at step edges (B area) showing increased moir\'e corrugation in STM as compared to G/Ir(111) (A and C areas) (70 $\times$ 70 nm$^2$; $U_T$ = 0.65 V; $I_T$ = 1.21 nA). Corresponding LEED image in the inset. (b) Areas with Ni intercalated underneath graphene (B areas) show reduced mean apparent height in the line profiles and the histogram. The histogram shows the frequency of apparent height values appearing in the magnification depicted in (c) (black curve) and within areas on terrace A, B or C (yellow, orange and brown curves, respectively). (c)  Magnification of the dotted square in (a) (46 $\times$ 8.6 nm$^2$).}
\end{figure}

\begin{figure*}
\includegraphics[width=5in]{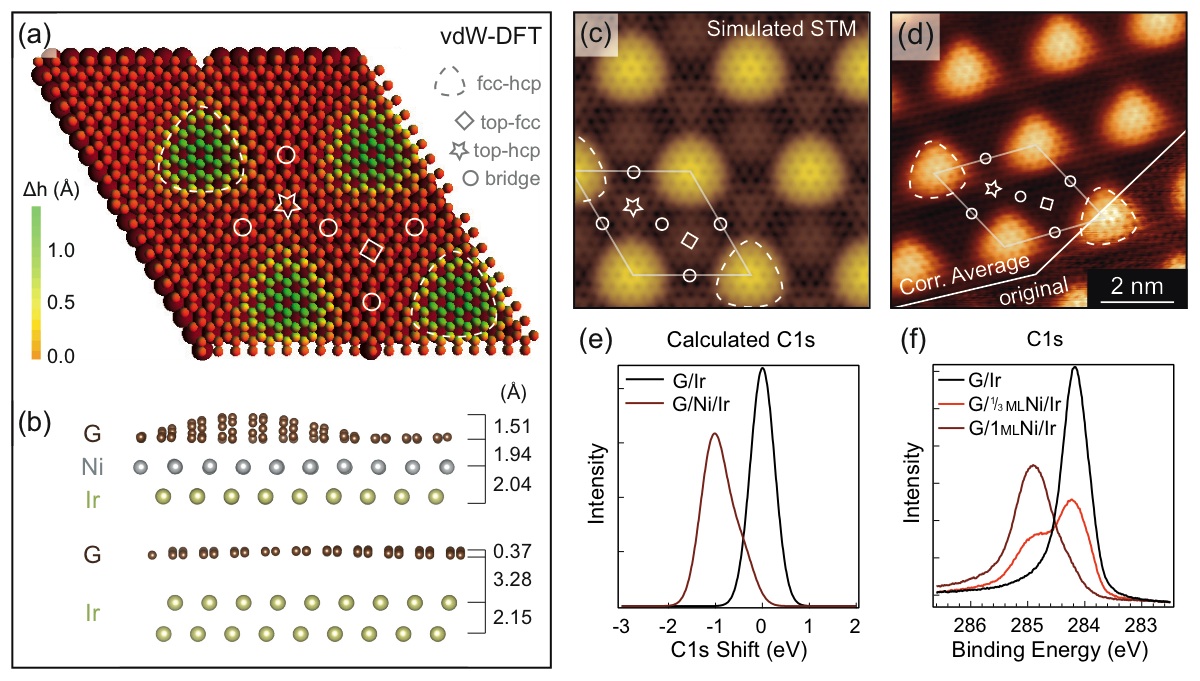}
\caption{\label{fig_stm_0:epsart} (color online) (a) Structural model for a single layer of graphene on Ni/Ir(111). The color coding indicates the height of the corrugation $\Delta h$ in the graphene layer.
(b) Comparison of the corrugation in the optimized structure of graphene/Ni/Ir(111) (upper panel) and graphene/Ir(111) (lower panel). (c) Simulated STM image for the states between E$_{F}$ and 0.2\,eV. (d) Atomically resolved STM topography of graphene/Ni/Ir(111) (8 $\times$ 8 nm$^2$; $U_T$ = 50.0 mV; $I_T$ = 35.0 nA). Theoretical (e) and experimental (f) C1s core level shifts of G/Ni/Ir(111) compared to G/Ir(111).}
\end{figure*}

\section{Results and Discussion}
In STM, the G/Ir(111) surface displays large fully graphene covered terraces with several hundreds of nanometers width and straight steps following the direction of the graphene moir\'e, consisting of distinct fcc-hcp carbon configurations and virtually indistinguishable top-hollow (top-fcc and top-hcp) sites. G/Ir(111) was imaged here in the dark-atop-contrast \cite{Diaye}, where elevated fcc-hcp regions appear as black depressions in middle of bright rings in STM topographies. Upon Ni intercalation straight terraces become disrupted and irregularly extended by areas with inverted moir\'e contrast (white protrusions on dark background) as compared to pristine G/Ir(111)  steps. Figure 1a depicts a detailed STM topography with graphene covering Ir (A, C) including an intermediate area with inverted contrast (B) and two remaining Ni clusters (in white) on top of graphene with height in the nanometer range. From the morphology of the sample after intercalation it becomes clear that areas A and C display pristine graphene on adjacent Ir substrate levels, whereas area B corresponds to graphene on a Ni-intercalated region. To shed more light on the overall as well as site specific graphene-substrate interaction we analyzed the area depicted in Figure 1c and evaluated line profiles across the terraces and histograms showing the distribution of apparent height values (Figure 1b).

Line profile 1 crosses the fcc-hcp sites of G/Ir(111) on terrace A, which appear as dark depressions, and continues across terrace B where bright protrusions now occupy the former fcc-hcp sites. Height profile 2 crosses the bright protrusions of terrace B. A large peak-to-valley height variation of 0.6\,\AA\ is measured on terrace B as compared to the much shallower peak-to-valley corrugation of 0.25\,\AA\ of G/Ir(111) on terraces A and C. In the histogram in Figure 1b the frequency of apparent height values is shown for equally sized areas on terrace A, B and C, respectively, as well as for the complete area in Figure 1c. For G/Ir(111) on terrace A and C the distribution is narrow (0.45 \AA\ peak width) and features a distribution maximum reflecting the top-hollow sites and a distinct shoulder at lower apparent height corresponding to the fcc-hcp regions. For Ni intercalated graphene on terrace B the distribution is much wider (0.8 \AA\ peak width) with a maximum 0.6 \AA\  below the maximum of G/Ir(111) and a shoulder extending far into the G/Ir(111) region. Assuming the intercalated Ni atoms arrange pseudomorphically on the Ir(111) surface with comparable interplane distance, the measured distance between equivalent points on the terraces A and B reflects to a large extent the difference in the graphene-metal distance (see also the discussion of the DFT results), and leads to the intriguing result that graphene on intercalated Ni is in average 0.6\,\AA\ closer to the topmost substrate layer compared to the pristine G/Ir(111). The almost unaffected continuation of the graphene moir\'e on intercalated Ni -- albeit with increased corrugation -- justifies the assumption of pseudomorphic arrangement of the intercalated Ni atoms. Moreover, the experimental data do not show any indication for the formation of a surface alloy.

The DFT calculations allow to investigate the structural changes in the  ($10 \times 10$) graphene sheet  adsorbed on a   ($9 \times 9$) Ni/Ir(111) substrate [1 ML Ni pseudomorphically arranged on Ir(111)]. It should be noted that the strength of the hybridization is closely related to the minimal graphene-substrate distances  \cite{Mittendorfer}. Consequently the structural analysis allows to deconvolute the effects of the lattice mismatch between the graphene sheet and the substrate, and the chemical properties of the interface. Figures 2(a-b) display the geometry of the graphene sheet after the relaxation. The model shows that the intercalation of the Ni layer leads to a pronounced corrugation of $\Delta h$ = 1.51 \AA~ in  the graphene layer, significantly larger than for G/Ir(111) (Figure 2b). Yet it should be noted that
more than 70\% of the carbon atoms in the graphene layer are adsorbed at a close distance of about 2.0 - 2.2 \AA~ from the Ni layer. The comparison with 
the inter-plane distance of G on  Ni(111) (2.1 \AA) therefore hints at a similar binding of graphene to the substrate, despite the large experimental strain of roughly 9\% due to pseudomorphic growth of the Ni lattice in G/1 ML Ni/Ir(111). Therefore, the G-Ni bonding seems to be mainly affected by the electronic contribution, which also drives the strong corrugation of the graphene layer in the mismatched structure. 

In the flat regions, the magnetic moment of the Ni atoms is completely quenched by the interaction with the graphene sheet, while the Ni atoms under the graphene bubbles yield a small magnetic moment ($<$0.4 $\mu_B$). We find that no magnetic moment is induced in the graphene sheet that can be due to the small net magnetic moment of the underlying Ni film, opposite to the graphene/Ni(111) system showing an induced magnetic moment of carbon atoms~\cite{Weser:2010}.

Seen from an atomistic point of view, the close adsorption configuration of the graphene layer is reached not only for the top-fcc configuration preferred on Ni(111), but also for the adsorption in a bridge-like configuration. Both configurations yield a close adsorption distance reaching values as low as 1.94 \AA. In contrast, the weak interaction in the fcc-hcp sites (green regions in 
Figure 2a) leads to the formation of local protrusions, with a maximal distance of 3.45 \AA~ to the Ni layer common for physisorbed graphene. Nevertheless, this distance is still smaller than the calculated maximal (vdW-DF) distance 3.7 \AA ~($\Delta h$ = 0.37 \AA) for the adsorption of G on the bare Ir(111) surface (Figure 2b). 

A direct comparison of the obtained STM data (Figure 2d) and a simulated STM image (Figure 2c) reflects the structure of the adsorbed graphene sheet: the elevated fcc-hcp regions appear brightest, while the low-lying areas with top-hollow configuration appear as a dark background. In agreement with the structure recently reported for G/Ru(0001)\cite{Iannuzzi} and G/Rh(111) \cite{Wang2, Sicot, Voloshina}, the regions where the graphene sheet is adsorbed in a local bridge configuration is the area of the smallest distance to the surface. These areas appear as faintly visible depressions in STM topographies. A peak-to-valley corrugation of up to 1 \AA\ fits well the corrugation of 1.3 \AA\ in the simulated image. On the atomic level, G/Ni/Ir(111) shows rings of carbon atoms everywhere within the moir\'e supercell in Figure 1d, however with the strongly bound areas a difference in the intensity between the neighboring atoms is observed indicating a broken sublattice symmetry.

Previous studies demonstrate that the corrugation and hybridization of the graphene layer with the metallic substrate  is strongly reflected in the C1s line-shape \cite{Preobrajenski, Miniussi}.
Figure 2f shows the C1s core level (CL) taken at 445 eV of G/Ir(111) and its evolution during the intercalation of about one third of monolayer and one monolayer of Ni atoms.  According to the existing literature \cite{Preobrajenski, Busse}, in the G/Ir(111) system the C1s binding energy is found at (284.10$\pm$0.20)eV. After the intercalation of 0.33 ML of Ni 
atoms a second component at higher binding energy is seen. The relative intensities of the two components is fully inverted for 1ML of Ni atoms intercalated. The main peak centered at 284.90 ($\pm$0.20) eV is close to the value found for graphene grown on bare Ni(111) \cite{Gruneis}, where a single peak at 284.7 ($\pm$0.18) eV was measured, with an intrinsic line width of 216 meV. In our system, the C1s line-shape exhibits a total width of about 840 meV and a strong asymmetry towards lower binding energy, likely convoluting different components. 
This is confirmed by the  calculated core level states of G/Ni/Ir(111): although the calculations predict only a minor CL shift for the carbon atoms in the elevated regions of the moir\'e pattern, the strongly interacting C atoms in the lower regions are dominant in the convolution considering all contributions (Figure 2e) and thus do not exhibit a double C1s peak as observed for G/Re(0001), G/Rh(111) or G/Ru(0001) \cite{Miniussi,Preobrajenski}.

\begin{figure*}
\includegraphics[width=6.5in]{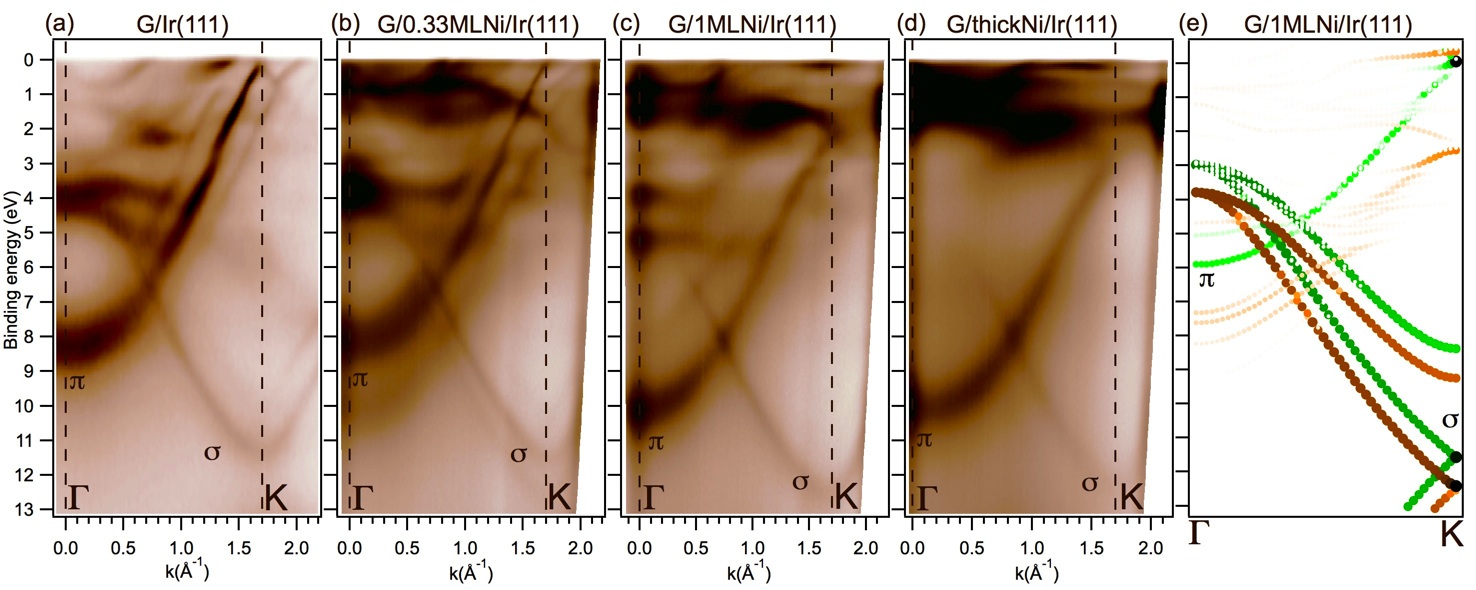}
\caption{\label{fig2:epsart} (color online) (a-d) ARPES dispersions along ${\Gamma}{\mbox{K}}$ as a function of the amount of Ni atoms intercalated underneath G/Ir(111). (a) and (d) show extreme cases of 0 ML and thick Ni, respectively. (e) Carbon projected band structure of a (1x1) model system of graphene/Ni/Ir(111). The band structure was evaluated at a G-Ni distance of 2.0 \AA~ (brown-dark gray dots) and 3.4 \AA~ (green-light gray dots).}
\end{figure*}

To shed more light on the overall interaction of graphene with the mismatched Ni layers, we have performed ARPES measurements mainly looking at the graphene $\pi$-band. Figures 3(a-d) show ARPES maps of the electronic band dispersion of (a) G/Ir(111); (b) G/0.33MLNi/Ir(111); (c) G/1MLNi/Ir(111); (d) G/thickNi/Ir(111), all taken along the $\Gamma$K direction. The last measurement was collected after the intercalation of several monolayers of Ni. The $\pi$ state of G/Ir(111) (Figure 3a) exhibits a minimum at the $\Gamma$-point at 8.30 eV, and approaches the Fermi level at the K-point at 70\,meV, where the planar $\sigma$ state reaches 11.35 eV of binding energy \cite{Rusponi}. According to the literature \cite{Preobrajenski, Rusponi, Starodub, Kralj, Papagno}, replicas bands of both $\pi$ and $\sigma$ states due to the moir\'e superpotential are seen close to the K-point. After the intercalation of about 1/3 ML of Ni atoms (Figure \ref{fig2:epsart}b), new $\pi$ and $\sigma$ states appear at higher binding energy together with the $\textit{d}$ states of Ni, weakly dispersing in the range 0-2\,eV below the Fermi level. The co-existence of double $\pi$ and $\sigma$ states reflects a non-homogenous surface with clean areas of G/Ir(111) and patches where Ni atoms are in between. When a full monolayer of Ni atoms is intercalated via annealing (Figure 3c),  the new band structure  evolves clearly. The $\pi$ state now exhibits a minimum at the $\Gamma$-point at 10.03 eV and reaches a maximum at about 2.16 eV at the K-point, where it merges with the $\textit{d}$ states of Ni. The $\sigma$ state is also shifted to a higher binding energy compared to G/Ir(111), with a minimum at the K-point at about 12.48 eV. When several monolayers of Ni (above 5) are intercalated via annealing (Figure 3d), the electronic states of Ir are no longer detected, and the band structure reflects that of G/Ni(111) \cite{Dedkov1}. Notably, while the $\sigma$ state is unaffected by the number of Ni layers at the K-point, the $\pi$ state exhibits a clear shift towards lower binding energies (up to 2.16 eV for 1 ML intercalated Ni) with respect to G grown on multilayers (Figure 3d) or bulk nickel \cite{Dedkov1}, where the $\pi$ band maximum is found at about 2.65 eV. This finding is related to the differences in width of the Ni-$\textit{d}$ states of a single intercalated Ni layer compared to a surface of bulk nickel: narrowing of the Ni\,3$d$ band upon the decreasing of the Ni layer thickness. Taking into account that the position of the graphene-derived $\pi$ band at the $\Gamma$ point is the same for both thick and thin (1\,ML) intercalated Ni layers, we can conclude that the energy shift of $\pi$ band with respect to free-standing graphene is purely defined by the charge transfer between Ni and C atoms at the closest distance through the donation/back-donation mechanism \cite{Kozlov}. At the same time the presence as well as the width of the band gap between $\pi$ and $\pi^*$ graphene-derived states is determined by the broken symmetry for two carbon atoms in the graphene unit cell in this system accompanied by a strong hybridization between Ni 3$d$ and graphene $\pi$ states.

For a comparison with experiment, we also evaluated the theoretical band structure. In order to avoid the back-folding induced by the larger supercell, we have mimicked the local interactions by calculating the electronic structure for a smaller (1$\times$1) model of G/Ni/Ir(111) in a top-fcc configuration at an average distance of the flat regions (2.0 \AA) and at the maximal height of the bubbles (3.4 \AA).
The resulting band structure is shown in Figure 3e. Although the effect of the lattice mismatch between the graphene sheet and the substrate is lost in this smaller model, the calculations clearly show that the interactions at the elevated regions of the bubbles (green-light gray dots in  Figure 3e)
are rather weak, resulting in a nearly unperturbed graphene band structure. On the other hand, a much stronger interaction can be expected for the dominant flat regions in the vicinity of the surface,
leading to a large splitting of the $\pi$ band at the Dirac point  (brown-dark gray dots in  Figure 3e). These findings agree with the experimentally observed opening of a band gap in the ARPES data. Furthermore, no $\pi$ band splitting is observed experimentally in our system, due to the metallic nature of graphene \cite{Brugger}, in contrast to the electronic behavior of $\textit{h}$-BN grown on selected transition metals \cite{Greber}, where the dielectric nature of the overlayer allows to observe double $\sigma$ and $\pi$ states corresponding to upper and lower regions.

\section{Conclusion}
In conclusion, we have shown that the adsorption of graphene on epitaxial layers allows to study the influence of the 
lattice mismatch between the graphene layer and the support, while keeping the chemical environment similar.   
For graphene on Ni/Ir(111), we find that the interaction is {\it locally} strongly enhanced for specific adsorption configurations.
Consequently, in contrast to  G/Ir(111), the moir\'e structure of G/Ni/Ir(111) exhibits a strong corrugation, with a modulation of about 1.5 \AA~and a minimum G-Me distance slightly smaller than 2 \AA~. 
The graphene band structure probed by ARPES shows a clear transition from a nearly-free standing to a strongly-hybridized character of the $\pi$ band, 
in analogy with graphene grown on bulk nickel.
The hybridization between Ni $\textit{d}$ states and graphene $\pi$ states is directly related to the strongly interacting top-hollow and bridge configurations in 
the lower parts of the moir\'e mesh, at a distance of about 2.0 - 2.2 \AA~ from the Ni layer.
In contrast, the interaction is significantly weaker for other regions (fcc-hcp configurations) of the moir\'e mesh, where only a van der Waals like  binding 
is observed.  
Therefore we can identify the role of two important contributions to the adsorption: while the electronic interaction  
dominates in the strongly interacting regions, the lattice mismatch between graphene and the metal support 
is decisive for the ratio between strongly and weakly interacting regions. We expect that this interplay is one of the key features 
for mismatched graphene-metal interactions. 

\section*{Acknowledgements}
This work has been supported by the Italian Ministry of Education, University, and Research (FIRB Futuro in Ricerca, PLASMOGRAPH Project), by the European Science Foundation (ESF) under the EUROCORES Program EuroGRAPHENE, by the Austrian Science Fund (FWF) project I422-N16, and by the Research Center
``UltraQuantum'' of the University of Konstanz (Excellence Initiative). The Vienna Scientific Cluster (VSC) is acknowledged for CPU time.


\begin{thebibliography}{999}

\bibitem{Rusponi} S. Rusponi, M. Papagno, P. Moras, S. Vlaic, M. Etzkorn, P. M. Shverdyaeva, D. Pacil\'e, H. Brune, and C. Carbone, Phys. Rev. Lett. {\bf 105}, 246803 (2010).

\bibitem{Nagashima} A. Nagashima, N. Tejima, and C. Oshima, Phys. Rev. B {\bf 50}, 17487 (1994).

\bibitem{Papagno} M. Papagno, S. Rusponi, P. M. Sheverdyaeva, S. Vlaic, M. Etzkorn, D. Pacil\'e, P. Moras, C. Carbone, and H. Brune, ACS Nano {\bf 6}, 199 (2012).

\bibitem{Varykhalov} A. Varykhalov, J. S\'anchez-Barriga, A. M. Shikin, C. Biswas, E. Vescovo, A. Rybkin, D. Marchenko, and O. Rader, Phys. Rev. Lett. {\bf 101}, 157601 (2008).

\bibitem{Batzill} M. Batzill, Surf. Sci. Rep. {\bf 67}, 83 (2012).

\bibitem{Wintterlin} J. Wintterlin and M.-L. Bocquet, Surf. Sci. {\bf 603}, 1841 (2009).

\bibitem{Dedkov:2012} Yu. S. Dedkov, K. Horn, A. Preobrajenskij, and M. Fonin, in Graphene Nanoelectronics (Springer, Berlin 2012).

\bibitem{Voloshina:2012} E. Voloshina and Yu. Dedkov, Phys. Chem. Chem. Phys. {\bf 14}, 13502 (2012).

\bibitem{Kozlov} S. M. Kozlov, F. Vi\~{n}es, and A. G\"orling, J. Phys. Chem. C {\bf 116}, 7360 (2012).

\bibitem{Hammer} B. Hammer and J. K. N$\o$rskov, Adv. Catal. {\bf 45}, 71 (2000). 

\bibitem{Mittendorfer} F. Mittendorfer, A. Garhofer, J. Redinger, J. Klime\v{s}, J. Harl, and G. Kresse, Phys. Rev. B  {\bf 84}, 201401(R) (2011).

\bibitem{Dedkov3} E. N. Voloshina, A. Generalov, M. Weser, S. B$\ddot{o}$ttcher, K. Horn, and Yu. S. Dedkov,  New J. Phys. {\bf 13}, 113028 (2011).

\bibitem{Adamska} L. Adamska, Y. Lin, A. J. Ross, M. Batzill, and I. I. Oleynik, Phys. Rev. B  {\bf 85}, 195443 (2012).

\bibitem{Gamo} Y. Gamo, A, Nagashima, M. Wakabayashi, M. Terai, and C. Oshima, Surf. Sci.  {\bf 374}, 61 (1997).  

\bibitem{Dedkov1} Yu. S. Dedkov and M. Fonin, New J. Phys. {\bf 12}, 125004 (2010).

\bibitem{vasp1} G. Kresse and J. Furthm\"uller, Comput. Mater. Sci. {\bf 6}, 15 (1996).

\bibitem{vasp2} G. Kresse and J. Hafner, Phys. Rev. B {\bf 47}, 558 (1993).

\bibitem{paw1} P.E. Bl\"ochl, Phys. Rev. B {\bf 50}, 17953 (1994).

\bibitem{paw2} G. Kresse and D. Joubert, Phys. Rev. B {\bf 59}, 1758 (1999).

\bibitem{vdW} J. Klime\u{s}, D. R. Bowler, and A. Michaelides, J. Phys.: Condens. Matter. {\bf 22}, 022201 (2010).

\bibitem{vdW2} J. Klime\u{s}, D. R. Bowler, and A. Michaelides, Phys. Rev. B {\bf 83}, 195131 (2011).

\bibitem{TersoffHamann} J. Tersoff and D. R. Hamann, Phys. Rev. B {\bf 31}, 805 (1985).

\bibitem{Diaye} A. T. N'Diaye, J. Coraux, T. N. Plasa, C. Busse, and T. Michely, New J. Phys.  {\bf 10}, 043033 (2008).

\bibitem{Weser:2010} M. Weser, Y. Rehder, K. Horn, M. Sicot, M. Fonin, A. B. Preobrajenski, E. N. Voloshina, E. Goering, and Yu. S. Dedkov, Appl. Phys. Lett. {\bf 96}, 012504 (2010). 

\bibitem{Iannuzzi} M. Iannuzzi and J. Hutter, Surf. Sci  {\bf 605}, 1360 (2011).

\bibitem{Wang2} B. Wang, M. Caffio, C. Bromley, H. Fr\"uchtl, and R. Schaub, ACS Nano {\bf 4}, 5773 (2010).

\bibitem{Sicot} M. Sicot,  P. Leicht, A. Zusan, S. Bouvron, O. Zander, M. Weser, Yu. S. Dedkov, K. Horn, and M. Fonin, ACS Nano {\bf 6}, 151 (2012).

\bibitem{Voloshina} E. N. Voloshina, Yu. S. Dedkov, S Torbr\"ugge, A. Thissen, and M. Fonin, Appl. Phys. Lett. {\bf 100}, 241606 (2012). 

\bibitem{Preobrajenski} A. B. Preobrajenski, M. L. Ng, A. S. Vinogradov, and N. M$\dot{a}$rtensson, Phys. Rev. B  {\bf 78}, 073401 (2008).

\bibitem{Miniussi} E. Miniussi, M. Pozzo, A. Baraldi, E. Vesselli, R. R. Zhan, G. Comelli, T. O. Mente\c{s}, M. A. Ni\~{n}o, A. Locatelli, S. Lizzit, and D. Alf\`e, Phys. Rev. Lett. {\bf 106}, 216101 (2011).

\bibitem{Busse} C. Busse, P. Lazi$\acute{c}$, R. Djemour, J. Coraux, T. Gerber, N. Atodiresei, V. Caciuc, R. Brako, A. T. N'Diaye, S. Bl\"ugel, J. Zegenhagen, and T. Michely, Phys. Rev. Lett. {\bf 107}, 036101(2011).

\bibitem{Gruneis} A. Gr\"uneis, K. Kummer, and D. V. Vyalikh, New J. Phys. {\bf 11}, 073050 (2009).

\bibitem{Starodub} E. Starodub, A. Bostwick, L. Moreschini, S. Nie, F. El Gabaly, K. F. McCarty, and E. Rotenberg, Phys. Rev. B  {\bf 83}, 125428 (2011).

\bibitem{Kralj} M. Kralj, I. Pletikosi$\acute{c}$, M. Petrovi$\acute{c}$, P. Pervan, M. Milun, A. T. N'Diaye, C. Busse, T. Michely, J. Fujii, and I. Vobornik, Phys. Rev. B  {\bf 84}, 075427 (2011).

\bibitem{Brugger} T. Brugger, S. G\"unther, B. Wang, J. H. Dil, M.-L. Bocquet, J. Ostewalder, J. Wintterlin, and T. Greber, Phys. Rev. B {\bf 79}, 045407(2009).

\bibitem{Greber} T. Greber, M. Corso, and J. Ostewalder, Surf. Sci. {\bf 603}, 1373 (2009).







\end{thebibliography}
\end{document}